\documentclass[iop]{emulateapj}

\slugcomment{Accepted 2014 February 4 for publication in the Astronomical Journal}
\shorttitle{New Cool Brown Dwarf Discoveries with WISE}
\shortauthors{Cushing et al.}

\newcommand\teff{\mbox{$T_\mathrm{eff}$}}

\newcommand{\WISEzerofourtenFull}{\mbox{WISE J041022.71$+$150248.4}}
\newcommand{\WISEzerofourten}{\mbox{WISE J0410$+$1502}}

\newcommand{\WISEzeroninefourthreeFull}{\mbox{WISE J094305.98$+$360723.5}}
\newcommand{\WISEzeroninefourthree}{\mbox{WISE J0943$+$3607}}

\newcommand{\WISEtwozerozerozeroFull}{\mbox{WISE J200050.19$+$362950.1}}
\newcommand{\WISEtwozerozerozero}{\mbox{WISE J2000$+$3629}}

\newcommand{\WISEtwentytwozeronineFull}{\mbox{WISE J220905.73$+$271143.9}}
\newcommand{\WISEtwentytwozeronine}{\mbox{WISE J2209$+$2711}}

\newcommand{\UGPSzeroseventwotwoFull}{\mbox{UGPS J072227.51$-$054031.2}}

\newcommand{\WISEfifteenfortyoneFull}{\mbox{WISE J154151.65$-$225024.9}}

\newcommand{\WISEeighteentwentyeightFull}{\mbox{WISE J182831.08$+$265037.7}}

\newcommand{\WISE}{\textit{WISE}}

\begin{document}


\title{Three New Cool Brown Dwarfs Discovered with the
  \textit{Wide-field Infrared Survey Explorer} (\textit{WISE}) and an
  Improved Spectrum of the Y0 Dwarf WISE J041022.71$+$150248.4}


\author{Michael C. Cushing\altaffilmark{a}, 
  J. Davy Kirkpatrick\altaffilmark{b},
  Christopher R. Gelino\altaffilmark{b},
  Gregory N. Mace\altaffilmark{c}, 
  Michael F. Skrutskie\altaffilmark{d}, \&
  Andrew Gould\altaffilmark{e}}

\altaffiltext{a}{Department of Physics and Astronomy, The University of Toledo, 2801 West Bancroft Street, Toledo, OH 43606, USA}

\altaffiltext{b}{Infrared Processing and Analysis Center, California
   Institute of Technology, Pasadena, CA 91125}

\altaffiltext{c}{Department of Physics and Astronomy, UCLA, Los
   Angeles, CA 90095}

\altaffiltext{d}{Department of Astronomy, University of Virginia, Charlottesville, VA 22904}

\altaffiltext{e}{Department of Astronomy, Ohio State University, 140 W. 18th Ave., Columbus, OH 43210}

\begin{abstract}

  As part of a larger search of \textit{Wide-field Infrared Survey
    Explorer} (WISE) data for cool brown dwarfs with effective
  temperatures less than 1000 K, we present the discovery of three new
  cool brown dwarfs with spectral types later than T7.  Using
  low-resolution, near-infrared spectra obtained with the NASA Infrared
  Telescope Facility and the \textit{Hubble Space Telescope} we derive
  spectral types of T9.5 for \WISEzeroninefourthreeFull, T8 for
  \WISEtwozerozerozeroFull, and Y0: for \WISEtwentytwozeronineFull.  The
  identification of \WISEtwentytwozeronineFull\ as a Y dwarf brings the
  total number of spectroscopically confirmed Y dwarfs to seventeen.  In
  addition, we present an improved spectrum (i.e. higher signal-to-noise
  ratio) of the Y0 dwarf \WISEzerofourtenFull\ that confirms the Cushing
  et al.\ classification of Y0.  Spectrophotometric distance estimates
  place all three new brown dwarfs at distances less than 12~pc, with
  \WISEtwozerozerozeroFull\ lying at a distance of only 3.9--8.0~pc.
  Finally, we note that brown dwarfs like \WISEtwozerozerozeroFull\ that
  lie in or near the Galactic plane offer an exciting opportunity to
  measure their mass via astrometric microlensing.

\end{abstract}
\keywords{infrared: stars --- stars: low-mass, brown dwarfs --- stars:
  individual (\WISEzerofourtenFull, \WISEzeroninefourthreeFull, \\ \WISEtwozerozerozeroFull, \WISEtwentytwozeronineFull)}

\section{Introduction}

The launch of the \textit{Wide-field Infrared Survey Explorer}
\citep[\WISE;][]{2010AJ....140.1868W} in late 2009 ushered in a new era
in the study of brown dwarfs with spectral types later than T8
(effective temperatures (\teff) less than roughly 700~K).  Previous
searches in the field with the Two Micron All Sky Survey
\citep[2MASS;][]{2006AJ....131.1163S,2002ApJ...564..421B,2004AJ....127.2856B},
the UKIRT Infrared Deep Sky Survey
\citep[UKIDSS;][]{2007MNRAS.379.1599L,2008MNRAS.390..304P,2010MNRAS.406.1885B}
and the Canada France Brown Dwarf Survey
\citep[CFHBD;][]{2008A&A...484..469D,2011AJ....141..203A} uncovered ten
brown dwarfs with spectral types later than T8 but the survey
sensitivity limits combined with their intrinsic faintness at red
optical and near-infrared wavelengths \citep[$M_H$ $\gtrsim$
19~mag;][]{2013ApJ...762..119M} limited the ability of these surveys to
identify large numbers of them.

In contrast, \WISE\ surveyed the entire sky at four mid-infrared
wavelengths centered at 3.4 $\mu$m, 4.6 $\mu$m, 12 $\mu$m, and 22 $\mu$m
(hereafter denoted $W1$, $W2$, $W3$, and $W4$ respectively) that are
ideally suited to detect cool brown dwarfs because these bandpasses span
the peak of their spectral energy distributions.  In particular, the
$W1$ and $W2$ bands are centered on a deep CH$_4$ absorption band at 3.3
$\mu$m and a region relatively free of opacity at 4.7 $\mu$m where a
large fraction of the total flux of cool brown dwarfs emerges \citep[see
Figure 2 of][]{2011ApJ...726...30M}.  The resulting $W1$$-$$W2$ color is
very red which makes the identification of cool brown dwarf candidates
relatively straightforward.  To date, \WISE\ data have been used to
identify roughly two hundred brown dwarfs with spectral types later than
T6
\citep{2011ApJ...726...30M,2011ApJ...735..116B,2011A&A...532L...5S,2011ApJS..197...19K,2011ApJ...743...50C,2012ApJ...753..156K,2012ApJ...759...60T,2012ApJ...760..152L,2013ApJS..205....6M,2013ApJ...767L...1L,2013PASP..125..809T},
including the sixteen spectroscopically confirmed Y dwarfs previously
published
\citep{2011ApJ...743...50C,2012ApJ...753..156K,2012ApJ...759...60T,2012ApJ...758...57L,2013ApJ...776..128K}.

The detailed study of these cool brown dwarfs promises to shed light on
the physics of ultracool atmospheres and constrain brown dwarf formation
theories.  In particular, simulations by \citet{2004ApJS..155..191B}
suggest that the space density of cool brown dwarfs is very sensitive to
the underlying mass function, a potential constraint on the yet unknown
formation mechanism(s) of brown dwarfs \citep[see][for a review of the
current theories]{2013arXiv1302.3954S}.  However, brown dwarfs obey a
mass-luminosity-age relation so measuring the brown dwarf present-day
mass function is difficult given the ages of the vast majority of brown
dwarfs are unknown.  The underlying mass function can be constrained,
however, by first forward modeling the evolution of a population of
brown dwarfs over time
\citep[e.g.,][]{2004ApJS..155..191B,2005ApJ...625..385A,2006MNRAS.371.1722D}
and then comparing the observed space densities of brown dwarfs to the
predictions of the models.  Given the relative ease with which effective
temperatures can be estimated for cool brown dwarfs, the most common
technique has been to compare the space densities of the brown dwarfs as
a function of spectral type to model effective temperature distributions
\citep[e.g.,][]{2008MNRAS.390..304P,2010MNRAS.406.1885B,2012ApJ...753..156K,2013MNRAS.430.1171D,2013MNRAS.433..457B}.
Results to date suggest that if the mass function is modeled as dN/d$M$
$\propto$ $M^{-\alpha}$, then $-1$ $<$ $\alpha$ $<$ 0.

Implicit in this technique is the necessity of a complete,
volume-limited sample of brown dwarfs.  In this paper, we report three
new discoveries from our on-going program to construct just such a
volume-limited sample of cool brown dwarfs using data from
\textit{WISE}.  In addition, we present an improved $J$- and $H$-band
spectrum of the the Y0 dwarf \WISEzerofourtenFull\ (hereafter
\WISEzerofourten) located at a distance of roughly 4 pc
\citep{2013ApJ...762..119M}.

\section{Candidate Selection}

\begin{figure*}[h]
\centerline{\hbox{\includegraphics[width=6in,angle=0]{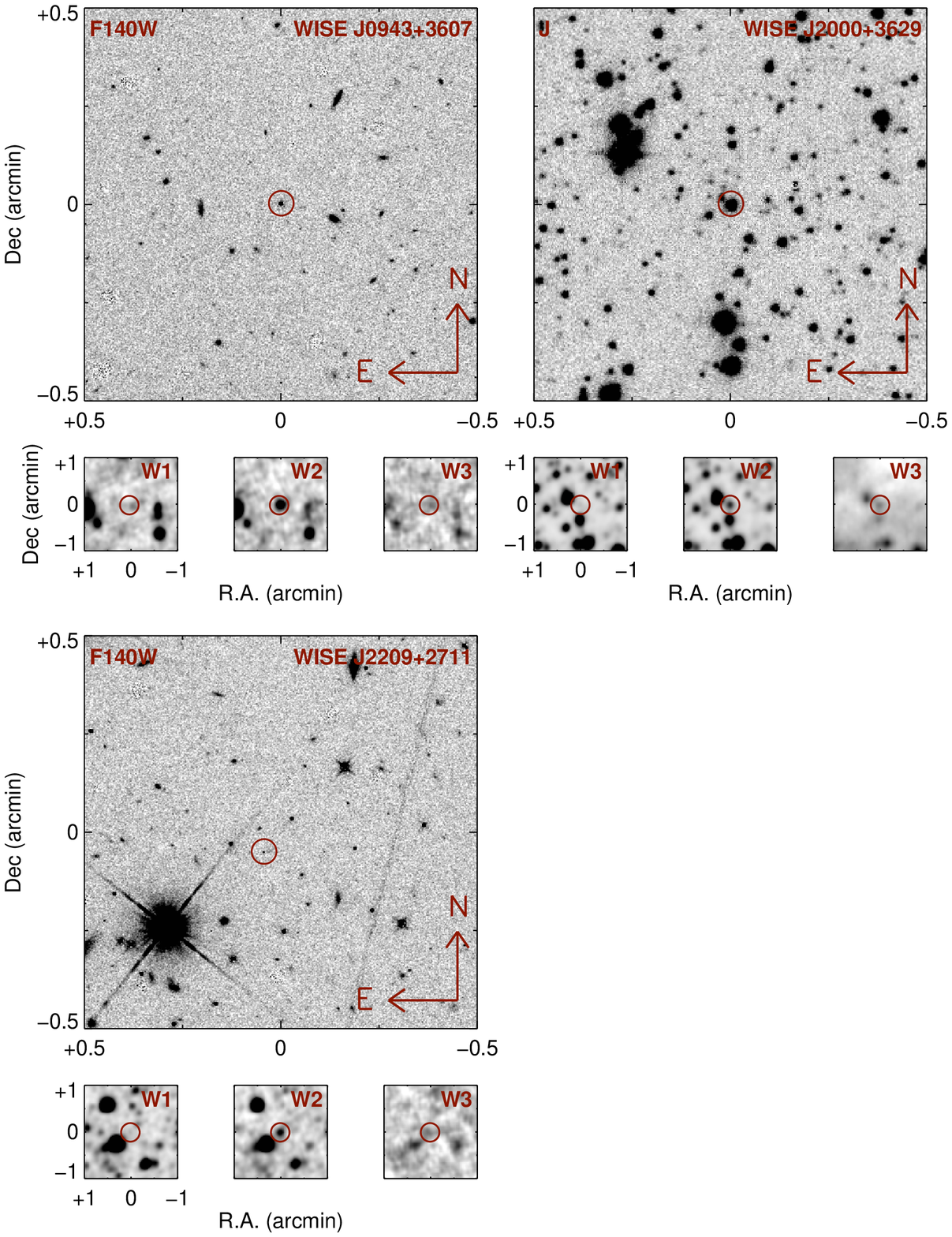}}}
\caption{\label{fig:finders}Finder charts for new brown dwarf
  discoveries, \WISEzeroninefourthree, \WISEtwozerozerozero, and
  \WISEtwentytwozeronine.  For each source, the lower three panels are 2
  $\times$ 2 arcmin \WISE\ $W1$, $W2$, and $W3$ band images while the
  larger panel is a 1 $\times$ 1 arcmin near-infrared image.  The
  near-infrared images for \WISEzeroninefourthree\ and
  \WISEtwentytwozeronine\ were obtained through the F140W filters of the
  WFC3 onboard \textit{HST} while the image for \WISEtwozerozerozero\ is
  from UKIDSS and was obtained through the MKO-NIR $J$-band filter.  All
  images are oriented such that North is up and East is to the left.}
\end{figure*}

\WISEzeroninefourthreeFull, \WISEtwozerozerozeroFull, and
\WISEtwentytwozeronineFull\ (hereafter \WISEzeroninefourthree,
\WISEtwozerozerozero\, and \WISEtwentytwozeronine) were identified as
candidate cold brown dwarfs as part of our larger search of the \WISE\
All-Sky Catalog and All-Sky Reject Table.  A description of the criteria
used to select brown dwarf candidates can be found in
\citet{2012ApJ...753..156K}.  Briefly, the primary selection criterion
is that the $W1-W2$ color from profile-fit photometry is greater than
2.0 mag which corresponds roughly to a spectral type of T6.  Table
\ref{tab:phot} lists the positions (as part of the \WISE\
identification) and \WISE\ photometry of the new brown dwarfs, and
Figure \ref{fig:finders} gives finder charts at both near- and
mid-infrared wavelengths.

\begin{turnpage}
\begin{deluxetable*}{rcccccccccc}
\tablecolumns{11}
\tabletypesize{\scriptsize} 
\tablewidth{0pc}
\tablecaption{\label{tab:phot}Photometry of New Brown Dwarfs}
\tablehead{
\colhead{} & 
\colhead{} & 
\colhead{} & 
\colhead{} & 
\multicolumn{4}{c}{\textit{WISE}\tablenotemark{a}} & 
\colhead{} & 
\multicolumn{2}{c}{\textit{Spitzer}} \\
\cline{5-8}
\cline{10-11}
\colhead{} \\
\colhead{Object} & 
\colhead{$J$} & 
\colhead{$H$} & 
\colhead{F140W\tablenotemark{b}} & 
\colhead{W1} & 
\colhead{W2} & 
\colhead{W3} & 
\colhead{W4} &
\colhead{} &  
\colhead{[3.6]} & 
\colhead{[4.5]} \\
\colhead{} & 
\colhead{(mag)} & 
\colhead{(mag)} & 
\colhead{(mag)} & 
\colhead{(mag)} & 
\colhead{(mag)} & 
\colhead{(mag)} & 
\colhead{(mag)} & 
\colhead{} & 
\colhead{(mag)} & 
\colhead{(mag)}}

\startdata

\WISEzeroninefourthreeFull & 19.74$\pm$0.05 & 20.37$\pm$0.20 & 20.04$\pm$0.02 & $>$17.99  & 14.37$\pm$0.06 & 12.33$\pm$0.33 & $>$8.89 & & 16.75$\pm$0.04 & 14.28$\pm$0.02 \\
\WISEtwozerozerozeroFull & 15.44$\pm$0.01  & 15.85$\pm$0.01  & $\cdots$ & 15.56$\pm$0.09 & 12.71$\pm$0.03 & 11.26$\pm$0.11 & $>$9.55 & & 14.22$\pm$0.02 & 12.68$\pm$0.02 \\
\WISEtwentytwozeronineFull & 22.58$\pm$0.14 &  22.98$\pm$0.31  & 23.16$\pm$0.16 & $>$18.47 & 14.79$\pm$0.07 & 12.44$\pm$0.34 & $>$9.31 & & 17.82$\pm$0.09 & 14.74$\pm$0.02 \\

\enddata

\tablecomments{All $J$- and $H$- band photometry is on the MKO-NIR
  system.}  \tablenotetext{a}{Magnitudes are from the WISE all-sky
  release, are on the Vega system, and are based on profile fits
  (w1mpro, w2mpro, w3mpro, w4mpro).  Upper limits are at the 95\%
  confidence level (see
  \url{http://wise2.ipac.caltech.edu/docs/release/allsky/expsup/sec4\_4c.html\#ul2}).}
\tablenotetext{b}{WFC3/\textit{HST} magnitudes are on the Vega system.}
\end{deluxetable*}
\end{turnpage}

\section{Observations}

We have an extensive and on-going observational followup campaign to
confirm brown dwarfs candidates identified using \WISE\ data.  In the
following sections, we describe the near- and mid-infrared photometric
and spectroscopic observations of \WISEzeroninefourthree,
\WISEzerofourten, \WISEtwozerozerozero, and \WISEtwentytwozeronine.  

\subsection{Near- and Mid-Infrared Imaging}

\subsubsection{UKIDSS}

\WISEtwozerozerozero\ was observed as part of the UKIDSS Galactic plane
Survey \citep{2008MNRAS.391..136L} in the Mauna Kea Observatories
Near-Infrared
\citep[MKO-NIR;][]{2002PASP..114..169S,2002PASP..114..180T,2005PASP..117..421T}
$J$ and $H$ bands on 2009-Jul-19 (UT).  The $J$- and $H$-band photometry
of \WISEtwozerozerozero\ (sourceID 438792120568) was obtained from Data
Release 7 using the WFCAM Science Archive and is given in Table
\ref{tab:phot}.

\subsubsection{WIRC/Hale 200-inch}

\WISEzeroninefourthree\ was targeted with the Wide-field Infrared Camera
\citep[WIRC;][]{2003SPIE.4841..451W} on the 200 inch Hale Telescope at
Palomar Observatory.  It was observed in the MKO-NIR $J$ and $H$ bands
on the night of 2010 Dec 24 (UT).  WIRC uses a 2048 $\times$ 2048
HAWAII-2 array with a plate scale of 0$\farcs$2487 pixel$^{-1}$
resulting in a 8$\farcm$7 $\times$ 8$\farcm$7 field of view.  Fifteen 60
sec exposures (900 sec total) were obtained in each of the MKO-NIR $J$
and $H$ bands.  A description of the data reduction can be found in
\citet{2011ApJS..197...19K}.  The $J$- and $H$-band magnitudes of
\WISEzeroninefourthree\ are given in Table 1.

\subsubsection{Keck II/NIRC2}

\WISEtwentytwozeronine\ was observed with NIRC2 behind the Keck II
LGS-AO system \citep{2006PASP..118..297W,2006PASP..118..310V} on the
night of 20 July 2011.  We used the $R$=13.3 USNO-B star 1171-0655438
\citep{2003AJ....125..984M} located 23$\arcsec$ from the target for the
tip-tilt reference star.  Images were obtained with the $J$ and $H$
MKO-NIR filters and the wide camera setting (nominal pixel scale =
0.039686$\arcsec$/pixel) which provided a single-frame field of view of
40\arcsec$\times$40\arcsec.  Twenty-seven $J$-band images (120~sec
exposure times) and fifty-four $H$-band images (60~sec exposure times)
were stacked and median averaged to create $J$ and $H$ mosaics with
total exposure times of 54 minutes.  The mosaics were photometrically
calibrated using two stars in common between the NIRC2 images and much
wider field-of-view, though shallower $J$ and $H$ WIRC images.

\subsubsection{WFC3/HST}

Four images each of \WISEzeroninefourthree\ and \WISEtwentytwozeronine\
through the F140W filter ($\lambda_p$=1392.3 nm, width=384 nm) were
obtained with the Wide Field Camera 3
\citep[WFC3;][]{2008SPIE.7010E..43K} on-board the \textit{Hubble Space
  Telescope} (HST) as a part of a Cycle 19 program to obtain grism
spectroscopy of Y dwarf candidates (GO-12544, PI=Cushing).  The WFC3
uses a 1024 $\times$ 1024 HgCdTe detector with a plate scale of
0$\farcs$13 pixel$^{-1}$ which results in a field of view of
123$\times$126 arcsecond.  Each exposure was 78 sec long for a total
exposure time of 312 sec.

For each source, the four images were combined using MULTIDRIZZLE
\citep{2002hstc.conf..337K}.  Photometry of each source was performed
using a 0$\farcs$4 radius aperture.  The noise in each pixel of a
drizzled image is correlated so the root-mean-squared noise in the local
background (i.e. a sky annulus) cannot be used to estimate the noise in
the background \citep[e.g.,][]{2000AJ....120.2747C}.  We therefore
measured the background in an aperture of 0$\farcs$4 at 1000 star-free
positions on the image.  The background level and uncertainty are given
by the mean and standard deviation of the resulting distribution.  The
F140W magnitudes of \WISEzeroninefourthree\ and \WISEtwentytwozeronine\
on the Vega system are given in Table \ref{tab:phot} and were computed
using the F140W zero point of
25.1845 mag\footnote{\url{http://www.stsci.edu/hst/wfc3/phot\_zp\_lbn}}.

\subsubsection{IRAC/\textit{Spitzer}}

\WISEzeroninefourthree, \WISEtwozerozerozero, and
\WISEtwentytwozeronine\ were observed with the Infrared Array Camera
\citep[IRAC;][]{2004ApJS..154...10F} on board the \textit{Spitzer Space
  Telescope} \citep{2004ApJS..154....1W} as part of Cycle 7 and Cycle 8
programs 70062 and 80109 (PI: Kirkpatrick).  We used the Infrared Array
Camera \citep[IRAC;][]{2004ApJS..154...10F} to image each dwarf in
bandpasses centered at 3.6 and 4.5 $\mu$m (hereafter denoted as [3.6]
and [4.5]).  These bandpasses are similar to the W1 and W2 bands
\cite[see Figure 2 of][]{2011ApJ...726...30M} and therefore provide more
sensitive measurements of the emergent flux at these wavelengths,
particularly at 3.6 $\mu$m where the WISE measurements are often upper
limits.  IRAC employs arrays with a pixel scale of 1$\farcs$2
pixel$^{-1}$ that cover a field of view of 5$\farcm$2 $\times$
5$\farcm$2.  A description of the data acquisition and reduction for our
\textit{Spitzer} programs can be found in \citet{2011ApJS..197...19K}.
The IRAC/\textit{Spitzer} [3.6] and [4.5] magnitudes of the three
sources are given in Table \ref{tab:phot}.  We note that
  \citet{2012AJ....144..148G} present photometry of
  \WISEzeroninefourthree\, and \WISEtwentytwozeronine\, using the same
  data but a slightly different algorithm and find similar magnitudes.

\subsection{Near-Infrared Spectroscopy}

\subsubsection{SpeX/IRTF}

A near-infrared spectrum of \WISEtwozerozerozero\ was obtained with SpeX
\citep{2003PASP..115..362R} on the 3.0 m NASA Infrared Telescope
Facility (IRTF) atop Mauna Kea, HI on the night of 2012 May 28 (UT).
The low-resolution prism mode was used in conjunction with a
0$\farcs$5-wide slit resulting in a spectrum that covered the wavelength
range 0.8$-$2.5 $\mu$m at a resolving power of $\sim$150.  A series of
ten, 120 sec exposures were obtained at two different positions along
the 15$''$-long slit for a total exposure time of 1200 sec.  The A0 V
star HD 192538 was observed for telluric correction and flux calibration
and argon arc and quartz-tungsten exposures for wavelength calibration
and flat fielding purposes.

\begin{figure} 
\centerline{\hbox{\includegraphics[width=3.5in,angle=0]{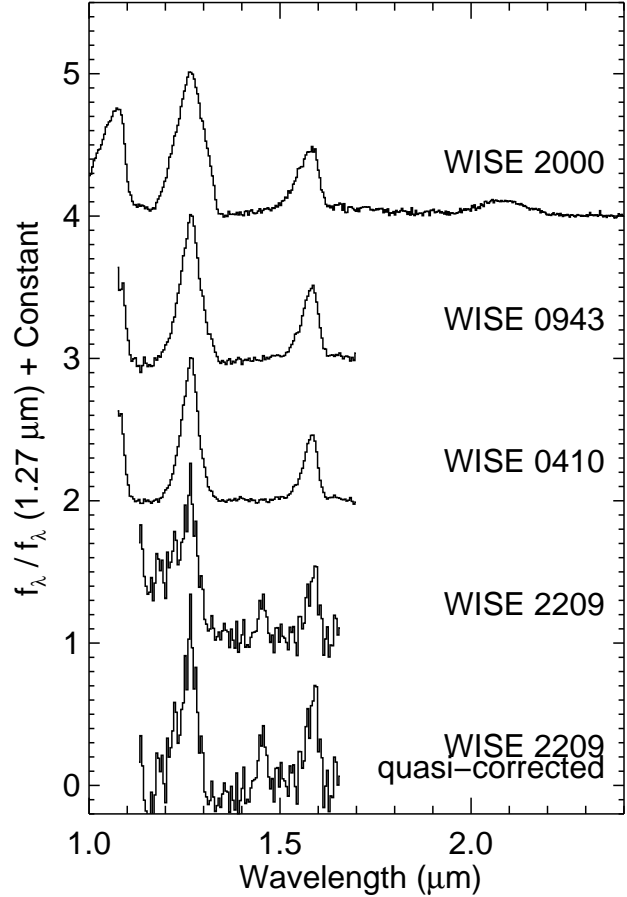}}}
\caption{\label{fig:spectra}Near-infrared spectra of
  \WISEtwozerozerozero, \WISEzeroninefourthree, \WISEzerofourten, and
  \WISEtwentytwozeronine.  The spectrum of \WISEtwentytwozeronine\ was
  contaminated so we subtracted a low-order polynomial as described in
  the text to produce a quasi-corrected spectrum (bottom spectrum).}
\end{figure}

The data were reduced using Spextool, the IDL-based data reduction
package for SpeX \citep{2004PASP..116..362C}.  Pairs of exposures taken
at the different slit positions were corrected for non-linearity,
pair-subtracted, and then flat-fielded.  The raw spectra were then
optimally extracted after subtracting any residual background and
wavelength calibrated using the argon arc spectrum.  The ten spectra
were then combined using a weighted average.  Telluric correction and
flux calibration were accomplished using the spectrum of HD 192538 and
the technique described by \citet{2003PASP..115..389V}.  The final
spectrum is shown in Figure \ref{fig:spectra} and has a signal-to-noise
ratio of 75, 110, 40, and 20 at the peaks of the $Y$, $J$, $H$, and $K$
bands, respectively.

\subsubsection{\label{sec:hst}WFC3/HST}

\WISEzerofourten, \WISEzeroninefourthree, and \WISEtwentytwozeronine\
were observed with the infrared channel of the Wide Field Camera 3
\citep[WFC3;][]{2008SPIE.7010E..43K} on-board the \textit{Hubble Space
  Telescope} (HST) as a part of a Cycle 19 program (GO-12544,
PI=Cushing).  The G141 grism was used to perform slitless spectroscopy
of each brown dwarf covering the 1.07$-$1.70 $\mu$m wavelength range at
a resolving power of $R$$\approx$130.  A description of the data
aquisition and reduction can be found in \citet{2011ApJ...743...50C}.
The final signal-to-noise ratio of the \WISEtwentytwozeronine\ spectrum
is low, with a peak of $\sim$6 in the $J$ band and $\sim$5 in the $H$
band while that of \WISEzeroninefourthree\ and \WISEzerofourten\ is
$\gtrsim$30 at the peaks of the $J$ and $H$ bands.

Spectroscopy with WFC3 is acheived via a slitless grism so some of the
resulting spectra could be contaminated by light from other stars in the
field.  Such is the case for \WISEtwentytwozeronine.  The contaminated
spectrum of \WISEtwentytwozeronine\ is shown in Figure \ref{fig:spectra}
and exhibits a smooth rise in flux towards shorter wavelengths due to
contaminating light of another star in the field.  We therefore
subtracted a second-order polynomial fit to the flux densities in the
1.1--1.2 $\mu$m, 1.3--1.5 $\mu$m, and 1.62--1.68 $\mu$m wavelength
intervals in order to produce a quasi-corrected spectrum which is also
shown in Figure \ref{fig:spectra}.

\section{Analysis}

\begin{figure*} 
\centerline{\hbox{\includegraphics[width=7in,angle=0]{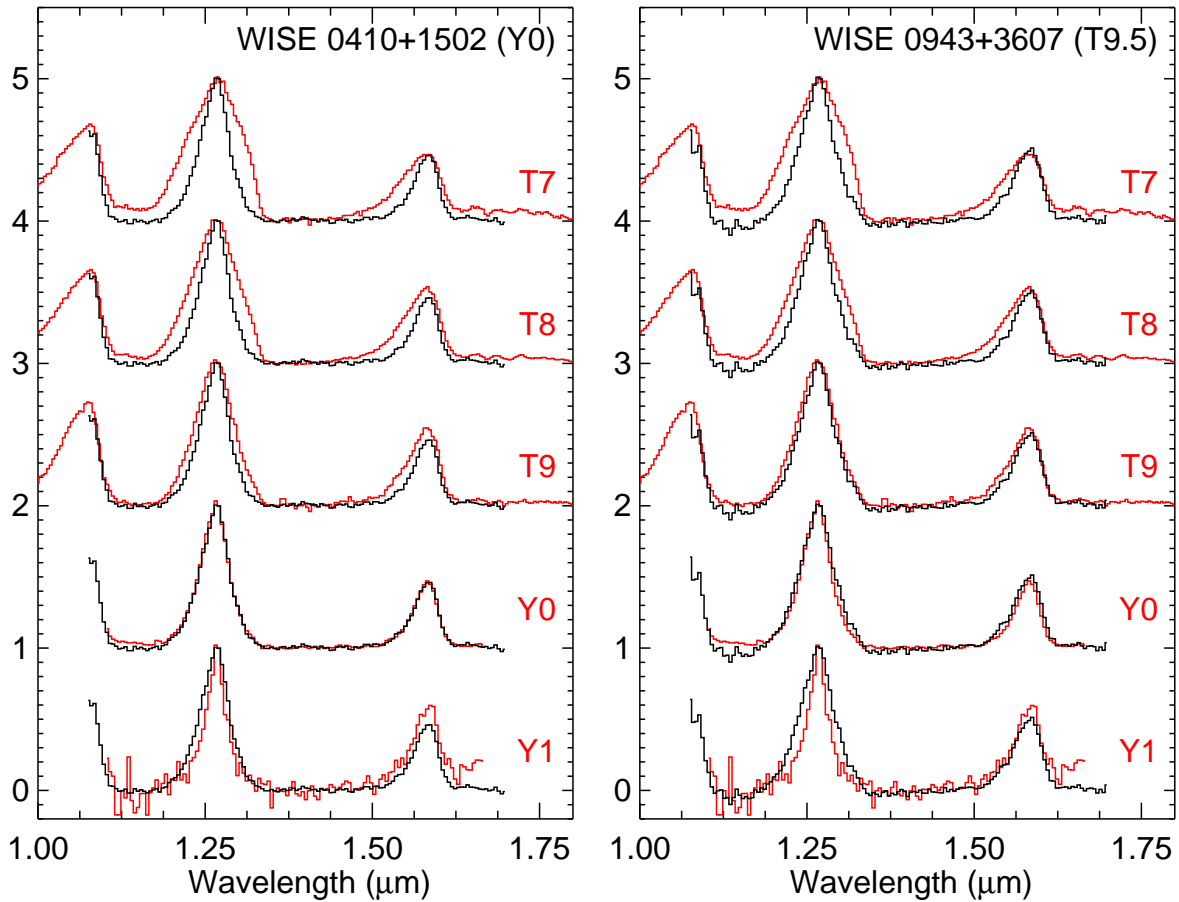}}}
\caption{\label{fig:sptypes1}Spectral classification of
  \WISEzerofourten\ and \WISEzeroninefourthree\  by
  direct spectral comparison.  The late-type T dwarf and tentative Y
  dwarf spectral standards from \citet{2006ApJ...637.1067B},
  \citet{2011ApJ...743...50C} and \citet{2012ApJ...753..156K} are shown
  in red while the spectra of the unclassified brown dwarfs are shown in
  black.}
\end{figure*}

\begin{figure*} 
\centerline{\hbox{\includegraphics[width=7in,angle=0]{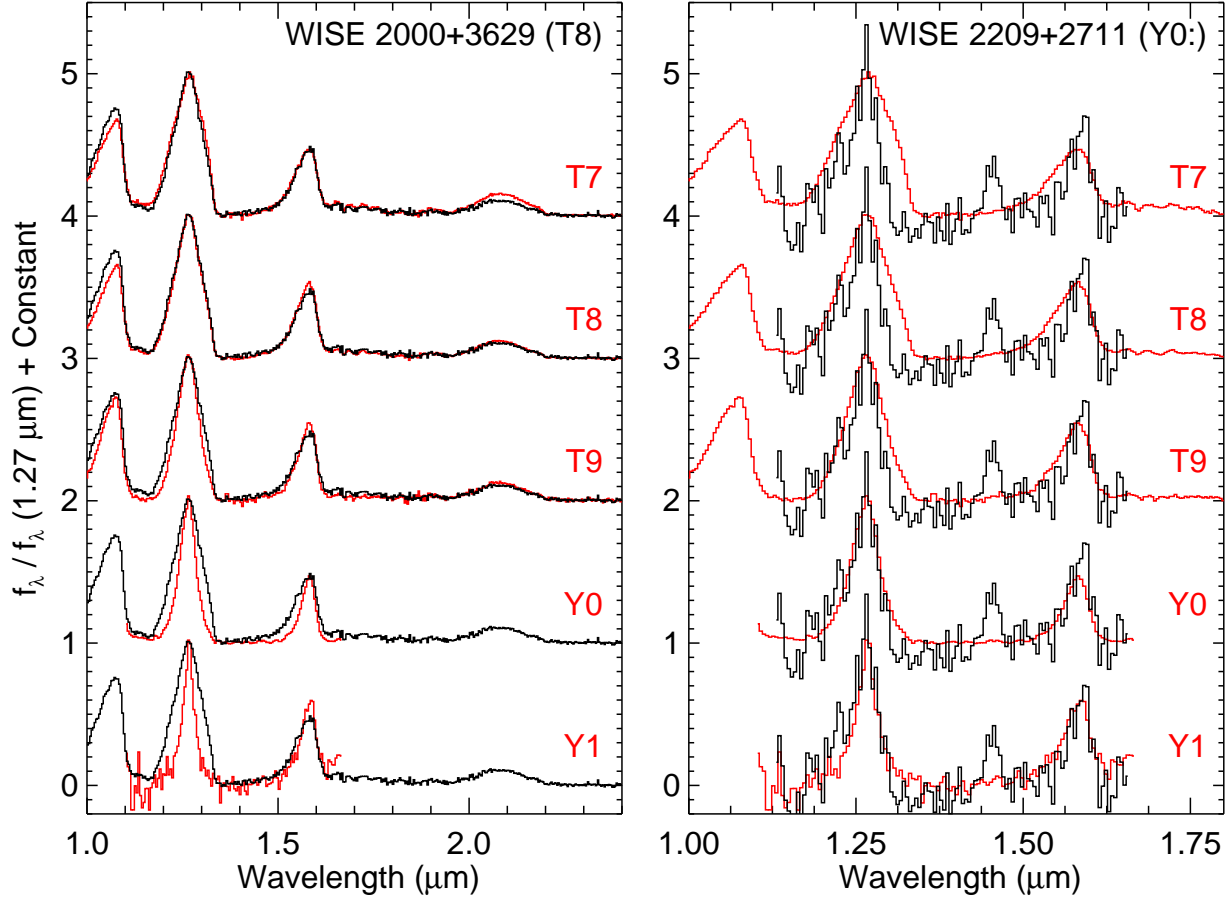}}}
\caption{\label{fig:sptypes2}Spectral classification of
  \WISEtwozerozerozero\ and \WISEtwentytwozeronine\ by direct spectral
  comparison.  The late-type T dwarf and tentative Y dwarf spectral
  standards from \citet{2006ApJ...637.1067B},
  \citet{2011ApJ...743...50C} and \citet{2012ApJ...753..156K} are shown
  in red while the spectra of the unclassified brown dwarfs are shown in
  black.}
\end{figure*}

The spectral types of \WISEzerofourten, \WISEzeroninefourthree,
\WISEtwozerozerozero, and \WISEtwentytwozeronine\ were determined by
direct comparison to the T and Y dwarf spectral standards of
\citet{2006ApJ...637.1067B}, \citet{2011ApJ...743...50C}, and
\citet{2012ApJ...753..156K}.  Figures \ref{fig:sptypes1} and
\ref{fig:sptypes2} shows the near-infrared spectra of each brown dwarf
plotted over the late-type T and Y dwarf spectral standards.  The final
spectral types are given in Table \ref{tab:properties} and a brief
discussion of some salient details regarding the classification of these
brown dwarfs is given below.

\WISEzerofourten\ is one of the six original Y dwarfs presented by
\citet{2011ApJ...743...50C}.  The \citeauthor{2011ApJ...743...50C}
spectrum obtained with the Folded-port InfraRed Echellette
\citep[FIRE;][]{2010SPIE.7735E..38S} spectrograph at the 6.5 m Magellan
Baade Telescope covers the 1.0--2.4 $\mu$m wavelength range at
$R$=250--350 but the signal-to-noise ratio was very low, reaching a peak
value of only $\sim$6 in the $J$ band. \WISEzerofourten\ was classified as Y0 based
primarily on the width of its $J$-band peak.  The higher signal-to-noise
ratio $HST$/WFC3 spectrum presented herein covers only the 1.10$-$1.65
$\mu$m wavelength range but confirms a spectral type of Y0 as it is a
near perfect match to the Y0 spectral standard WISE
J173835.53$+$273259.0 (left panel of Figure \ref{fig:sptypes1}).

The quasi-corrected spectrum of \WISEtwentytwozeronine\ is shown in the
right panel of Figure \ref{fig:sptypes2}.  Based on the width of its
$J$-band peak, \WISEtwentytwozeronine\ is clearly later than T9.  The Y0
spectral standard is a reasonable match but the low signal-to-noise
ratio of the spectrum makes it difficult to determine whether it could
be as late as Y1 so we classify \WISEtwentytwozeronine\ as Y0:.  An
uncontaminated and higher signal-to-noise ratio spectrum will be
required for a more precise spectral type.

Central to the construction of a volume-limited sample is accurate
knowledge of the distances to the nearby stars and brown dwarfs.  The
new brown dwarfs presented herein are currently being observed as part
of our program to measure parallaxes to the sample of cold brown dwarfs
in the solar neighborhood.  We can nevertheless derive
spectrophotometric distance estimates to them using the $H$ and $W2$
absolute magnitude spectral type relations of
\citet{2012ApJ...753..156K}\footnote{\citet{2013ApJ...762..119M}
  rederived both the $M_H$ and $W2$ versus spectral type relations with
  updated distances but found similar results to
  \citet{2012ApJ...753..156K}.}.  The range of possible distances for
each bandpass is given in Table \ref{tab:properties} and includes the
half subclass uncertainty in the spectral types and the photometric
uncertainties.  The $W2$ spectrophotometric distance estimates place the
three new brown dwarfs at distances between 6 and 12 pc.
\WISEtwozerozerozero\ lies at a distances of 3.9$-$7.3 pc based on its
$H$-band magnitude and a distance of 6.4$-$8.0 pc based on its $W2$
magnitude.  We therefore estimate its distance to be 3.9$-$8.0 pc which
places it within the 8 pc sample.

\section{Discussion}

\begin{figure} 
\centerline{\hbox{\includegraphics[width=3.5in,angle=0]{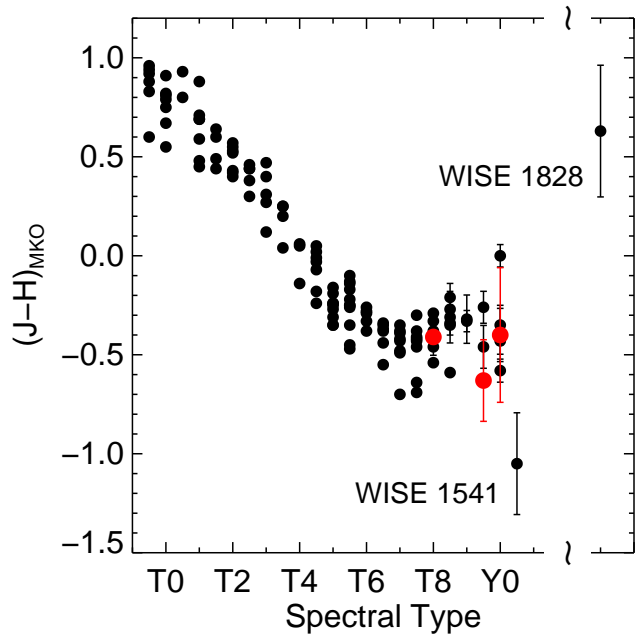}}}
\caption{\label{fig:nirvssptypes}MKO-NIR $J-H$ colors of T and Y dwarfs
  as a function of spectral type.  The filled black circles are from
  \citet{2010ApJ...710.1627L,2013ApJ...763..130L} while the red filled
  circles are from this work.  \WISEeighteentwentyeightFull\ is plotted
  separately because its spectral type is currently $\geq$Y2.  The three
  new brown dwarfs fall in line with the sequence suggesting they are
  not peculiar.}
\end{figure}

The classification of \WISEtwentytwozeronine\ as a Y dwarf brings the
total number of \textit{spectroscopically} confirmed Y dwarfs to
seventeen\footnote{WD 0806$-$661 \citep{2011ApJ...730L...9L} and CFBDSIR
  J1458$+$1013B \citep{2011ApJ...740..108L} have estimated effective
  temperatures similar to those of the known population of Y dwarfs but
  spectra of these cold brown dwarfs have yet to be obtained due to
  their faintness.}.  The colors of the late-type T dwarfs and Y dwarfs
exhibit distinct changes at the T/Y boundary with the $J-H$ colors
plateauing (albeit with a large scatter)
\citep{2011ApJ...743...50C,2012ApJ...753..156K} and the $z'-J$, $z'-H$,
and $Y-J$ colors rapidly becoming bluer
\citep{2012ApJ...758...57L,2013A&A...550L...2L}.  Figure
\ref{fig:nirvssptypes} shows the MKO $J-H$ colors of a sample of T and Y
dwarfs \citep{2010ApJ...710.1627L,2013ApJ...763..130L} as a function of
spectral type.  The $J-H$ colors plateau in the late-T spectral types
and turn redward at later types (with the exception of the Y0.5 dwarf
\WISEfifteenfortyoneFull).  The three new brown dwarfs are shown as
filled red circles and fall in line with the known sequence suggesting
that they are not peculiar.

The identification of three new late-type brown dwarfs that lie within
roughly 10 pc of the Sun is an important addition to the sample of
nearby cold brown dwarfs.  We do, however, defer updating the space
density estimates of the late-type T dwarfs and Y dwarfs presented in
until the parallaxes of these brown dwarfs have been measured and more
nearby cool brown dwarfs are identified.

One of the three new brown dwarfs, \WISEtwozerozerozero, is located in a
region ($l$=72$\fdg$7, $b$=$+$3$\fdg$32) covered by the UKIDSS Galactic
Plane Survey \citep{2008MNRAS.391..136L}.  \citet{2010MNRAS.408L..56L}
searched the Sixth Data Release of the Galactic Plane Survey to search
for late-type T or Y dwarf candidates but identified only a single
candidate - the well-known T9 dwarf \UGPSzeroseventwotwoFull.  The
position of \WISEtwozerozerozero\ was not covered by the Sixth Data
Release but even if it had been, \WISEtwozerozerozero\ would have been
missed by \citet{2010MNRAS.408L..56L} because its values of
\texttt{k\_1Ell} and \texttt{jppErrbits}, two flags in the UKIDSS
catalog that describe the source ellipticity and photometric quality,
fall outside of the \citeauthor{2010MNRAS.408L..56L} search criteria.
This suggests that perhaps more brown dwarfs in the Galactic plane await
discovery.

Indeed \WISEtwozerozerozero\ joins a growing list of brown dwarfs
identified in or near the Galactic plane including
\UGPSzeroseventwotwoFull\ and Luhman's binary system WISE
J104915.57$-$531906.1 \citep{2013ApJ...767L...1L}.  Previous searches
for brown dwarfs including our own ignore either part of, or the entire,
Galactic plane in order to minimize the impact of stellar contamination
\citep[e.g.,][]{2002ApJ...564..421B,2007AJ....133..439C}.  However, once
a brown dwarf has been discovered in the plane, the reasonably high
surface density of background stars offers us an opportunity to measure
the mass of the brown dwarf via astrometric microlensing
\citep{1964MNRAS.128..295R,1995AcA....45..345P,1996ApJ...470L.113M}.  

\begin{deluxetable}{lccc}
\tablecolumns{11}
\tabletypesize{\scriptsize} 
\tablewidth{0pc}
\tablecaption{\label{tab:properties}Derived Properties of New Brown Dwarfs}
\tablehead{
\colhead{Object} & 
\colhead{Spectral} & 
\colhead{$M_H$} & 
\colhead{$W2$} \\
\colhead{} & 
\colhead{Type} & 
\colhead{Distance} & 
\colhead{Distance}}

\startdata

\WISEzeroninefourthree\ & T9.5 & 7.9--34.3 & 6.6--12.1 \\
\WISEtwozerozerozero\ & T8 & 3.9--7.3 & 6.4--8.0 \\
\WISEtwentytwozeronine\ & Y0: & $\cdots$ & 5.2--11.7 \\

\enddata

\end{deluxetable}

During a microlensing event, the lens (i.e. the brown dwarf) both
magnifies the source (i.e. the background star) and shifts its centroid.
If the angular separation between the lens and the source is much larger
than the Einstein radius, then the magnification is negligible but the
centroid shift remains
\citep[e.g.,][]{1996ApJ...470L.113M,2001PASP..113..903G}.  If the lens
is also dark (i.e. is much fainter than the source) then a measurement
of the centroid shift of the source is relatively straightforward and
yields the mass of the lens.  Cool T and Y dwarfs are ideal targets for
astrometric microlensing measurements at optical wavelengths because
they are effectively dark at these wavelengths \citep[$M_V$ $\gtrsim$ 26
mag,][]{2010A&A...510A..99K,2013arXiv1312.1303B}.  The expected number
of lensing events per year, which is derived in the Appendix, is given
by:

\begin{equation}
N = 2\kappa \pi_L \mu_L \sigma_\mathrm{bg} \left ( \frac{\delta M_L}{\delta \Delta \theta} \right ),
\end{equation}

\noindent
where $\kappa \equiv \frac{4G}{c^2\, \mathrm{AU}} \approx 8.144\,
\frac{\mathrm{mas}}{M_\odot}$, $\pi_L$ is the parallax of the lens in
units of arcseconds, $\mu_L$ is the proper motion of the lens in units
of arcseconds per year, $\sigma_\mathrm{bg}$ is the surface density of
stars in units of stars per square arcseconds, $\delta M_L$ is the
required precision of the mass determination in solar masses, and
$\delta \Delta \theta$ is the precision of the astrometric measurement
in milliarcseconds.

Although we cannot compute a rate for \WISEtwozerozerozero\ because its
parallax and proper motion are unknown, we can get a sense of the
expected rate of brown dwarf lensing events near the Galactic plane by
considering a hypothetical brown dwarf located at $l=45^\circ$,
$b=0^\circ$, at a distance of 10 pc, and with a proper motion of
$0\farcs$5 yr$^{-1}$.  We used the Basan\c{c}on model of the Galaxy
\citep{2003A&A...409..523R} to generate a simulated population of stars
in a 0$\fdg$1 $\times$ 0$\fdg$1 field at this position which results in
a surface density of stars with $V<$ 24 of 0.12 stars per square
arcsecond.  If we require a mass precision of $\delta M_L$=0.10
$M_\odot$ and assume we can achieve an astrometric precision of $\delta
\Delta \theta$=0.2 mas (e.g., with the \textit{Hubble Space Telescope}),
then we find that $N\approx$ 0.05 or roughly one event every two
decades.  The potential to measure the masses of even a handful of cool
field brown dwarfs on a decades timescale should therefore stoke
interest in searches for brown dwarfs in the zone of avoidance.

Computing values for \UGPSzeroseventwotwoFull\ and Luhman's binary in a
similar fashion we find 0.02 and 0.3, respectively.  The value for
Luhman's binary is remarkably high, one event every few years, because
of its close distance ($\sim$ 2 pc) and high proper motion (2$\farcs$7
yr$^{-1}$).  Followup observations to determine whether it will actually
lens a nearby star should therefore be a high priority since a mass
measurement of such a nearby (binary) brown dwarf would be invaluable.

\acknowledgements
 
M.C.C. acknowledges fruitful discussions with Adam Schneider.  We thank
P. Eisenhardt for obtaining the NIRC2 data for \WISEzeroninefourthree.
Data presented herein were obtained at the W. M. Keck Observatory from
telescope time allocated to the National Aeronautics and Space
Administration through the agency's scientific partnership with the
California Institute of Technology and the University of California. The
Observatory was made possible by the generous financial support of the
W. M. Keck Foundation.  This publication makes use of data products from
the Wide-field Infrared Survey Explorer, which is a joint project of the
University of California, Los Angeles, and the Jet Propulsion
Laboratory/California Institute of Technology, funded by the National
Aeronautics and Space Administrations, the Two Micron All Sky Survey,
which is a joint project of the University of Massachusetts and the
Infrared Processing and Analysis Center/California Institute of
Technology, funded by the National Aeronautics and Space Administration
and the National Science Foundation and is based [in part] on
observations made with the \textit{Spitzer Space Telescope}, which is
operated by the Jet Propulsion Laboratory, California Institute of
Technology under a contract with NASA. Support for this work was
provided by NASA through an award issued by JPL/Caltech. and the
NASA/ESA Hubble Space Telescope, obtained at the Space Telescope Science
Institute, which is operated by the Association of Universities for
Research in Astronomy, Inc., under NASA contract NAS 5-26555. The HST
observations are associated with program \#12544, support for which was
provided by NASA through a grant from the Space Telescope Science
Institute, which is operated by the Association of Universities for
Research in Astronomy, Inc., under NASA contract NAS 5-26555.  This
research has made use of the NASA/IPAC Infrared Science Archive, which
is operated by the Jet Propulsion Laboratory, California Institute of
Technology, under contract with the National Aeronautics and Space
Administration.  M.C.C. was a visiting Astronomer at the Infrared
Telescope Facility, which is operated by the University of Hawaii under
Cooperative Agreement no. NNX-08AE38A with the National Aeronautics and
Space Administration, Science Mission Directorate, Planetary Astronomy
Program.  This research made use of the OSX Version of SCISOFT assembled
by Dr. Nor Pirzkal and F. Pierfederici.  A.~G. was supported in part by
NSF AST-1103471.

\appendix

The rate at which we can expect a nearby brown dwarf to astrometrically
lens a background star can be calculated as follows.  Let
$\theta_\mathrm{max}$ be the maximum projected angle separation between
the lens (i.e. the brown dwarf) and the source (i.e. a background star)
for which we can measure the mass of the lens to a fixed precision via
astrometric microlensing.  The number of such events per year is then
given by:

\begin{equation}
N = 2 \theta_\mathrm{max} \mu_\mathrm{L} \sigma_\mathrm{bg},
\end{equation}

\noindent
where $\mu_L$ is the proper motion of the lens in units of arcseconds
per year (and thus 2$\theta_\mathrm{max} \mu_L$ is the total area swept
out by the lens in a year) and $\sigma_\mathrm{bg}$ is the surface
density of background stars in units of stars per square arcsecond.  An
expression for $\theta_\mathrm{max}$ can be derived thusly.  Let
$\theta$ be the projected angular separation between the lens and the
source and $\theta_E$ be the angular Einstein radius given by:


\begin{equation}
\theta_\mathrm{E} = \sqrt{\frac{4GM_L}{c^2}\frac{D_S - D_L}{D_SD_L}}
\end{equation}
\noindent
where $M_L$ is the mass the lens, $D_S$ is the distance to the source,
$D_L$ is the distance to the lens, and the other constants have their
normal meaning.  In the limit that $\theta >> \theta_\mathrm{E}$, the
angular shift in the centroid of the source for a dark lens
(i.e. assuming the brown dwarf is much fainter than the source) is given
by:

\begin{equation}
\Delta \theta = \frac{\theta_\mathrm{E}^2}{\theta},
\end{equation}

\noindent
\citep{2000ApJ...534..213D}.  In the limit where $D_S >> D_L$, the
Einstein radius can be written as,

\begin{equation}
\theta_\mathrm{E} = \sqrt{\frac{4GM_L}{c^2D_L}},
\end{equation}

\noindent
and Equation 3 becomes,

\begin{equation}
\Delta \theta = \frac{4GM_L}{c^2 \theta D_L}.
\end{equation}

\noindent
Taking the differential of this equation holding $\theta$ and $D_L$
constant we have:

\begin{equation}
\delta \Delta \theta = \frac{4G\delta M_L}{c^2 \theta D_L}.
\end{equation}

\noindent
If we define $\theta_\mathrm{max}$ to be the maximum projected angular
separation between the lens and source given $\delta M_L$, the required
precision of the mass determination, and $\delta \Delta \theta$, the
precision of the astrometric measurement, we have:

\begin{equation}
\theta_\mathrm{max} = \frac{4G \delta M_L}{c^2 D_L \delta \Delta \theta}.
\end{equation}

\noindent
Define $\kappa$, a constant commonly used in the microlensing literature
\citep[e.g.,][]{2000ApJ...542..785G}, as

\begin{equation}
  \kappa \equiv \frac{4G}{c^2\, \mathrm{AU}} \approx 8.144\, \frac{\mathrm{mas}}{M_\odot},
\end{equation}

\noindent
and we can rewrite Equation 7 as

\begin{equation}
\theta_\mathrm{max} = \frac{\kappa \delta M_L \pi_L}{\delta \Delta \theta},
\end{equation}

\noindent
where $\pi_L$ is the parallax of the lens measured in arcseconds, and
$\delta M_L$ and $\delta \Delta \theta$ are now measured in solar masses
and milliarcseconds, respectively.  Substituting into Equation 1 we have
for the number of astrometric microlensing event per year:

\begin{equation}
N = 2\kappa \pi_L \mu_L \sigma_\mathrm{bg} \left ( \frac{\delta M_L}{\delta \Delta \theta} \right ).
\end{equation}

\bibliographystyle{apj}
\bibliography{ref,tmp}

\clearpage

\clearpage

\clearpage

%
%

%
%

\end{document}